\documentclass[aps,prb,twocolumn,showpacs,showkeys,amsmath,amssymb,superscriptaddress]{revtex4}
\usepackage{dcolumn}
\usepackage{bm}
\usepackage{graphicx}
\usepackage{epsfig}
\usepackage{amsmath}
\usepackage{natbib}
\usepackage{bm}

\begin{document}

\title{{\it Ab initio} study of helium and hydrogen interactions in $\alpha$-Fe}

\author{C. J. Ortiz}
\affiliation{Laboratorio Nacional de Fusi\'on por Confinamiento Magn\'etico,
CIEMAT, Avenida Complutense, 40, 28040, Madrid, Spain}
\date{\today}
\author{R. Vila}
\affiliation{Laboratorio Nacional de Fusi\'on por Confinamiento Magn\'etico,
CIEMAT, Avenida Complutense, 40, 28040, Madrid, Spain}
\author{J. M. Pruneda}
\email[]{miguel.pruneda@cin2.es}
\affiliation{Centre d'Investigaci\'on en Nanoci\'encia i Nanotecnolog{\'{\i}}a (CSIC-ICN). Campus de la UAB, E-08193 Bellaterra, Spain}
\date{\today}

\begin{abstract}
Density Functional Theory (DFT) calculations show a weak interaction between hydrogen and helium in iron, 
in contrast to previous reports of a strong trapping of hydrogen at helium\cite{Jiang2010,Tian2010}.  
The strong preference of He and H to occupy regions with low electronic density (such as vacancies) 
explains this discrepancy, with vacancy-He and vacancy-H binding forces concealing the repulsive interaction 
between He and H. Furthermore, Rate Theory simulations based on our DFT-calculated V$_n$He$_m$H$_p$ cluster 
energetics predict, as it is observed in some experiments, that synergetic effects could be expected between 
H and He in iron under irradiation.
\end{abstract}

\pacs{73.20.-r, 73.20.Hb, 73.22.Pr}

\maketitle

Hydrogen and helium are responsible for more than 98\% of nuclear matter in today's universe.
Understanding how they interact is critical for fundamental fields of research such as modeling 
the evolution of stars,~\cite{Fallon1960} White Dwarfs and the interior of 
Jovian planets,~\cite{Morales2009,Lorenzen2009}; but also in more technological applications such 
as sequential H/He implantation in the semiconductor industry,~\cite{Agarwal1998,Nguyen2007}, 
or to predict the agglomeration of defects and H and He that arise by nuclear transmutation between 
high energy neutrons and lattice atoms in materials irradiated in fusion conditions. 
In this sense, substantial research efforts have been devoted to the experimental and theoretical study of the 
interaction of He-defects,~\cite{Ortiz2007_1,Ortiz2009_1,Fu2004,Fu2005_2,Heinisch2006_1,Heinisch2007_1,
Vassen1991,Gao2006_1,Morishita2006_1, Sugano2002_1,Caro2011_1,Jourdan2011_1,Xu2009_1,Xu2010_1,
Caturla2007_1} or H-defects\cite{Counts2010,Tateyama2003} in iron-based alloys and different 
ferritic steels which are important structural materials for fusion reactors.
%
%
A few experimental investigations evidenced that helium could act as trap for hydrogen, hence reducing 
diffusivity and increasing embrittlement in iron alloys,~\cite{Schroeder91,Fukushima85, Tanaka2004_1} 
although experiments could not clarify whether the origin of these synergetic effects was due to the 
interaction between H and He itself or due to the trapping at damage produced during irradiation. 
A detailed study of interactions between hydrogen, helium and displacement damage is thus necessary to 
understand and predict their simultaneous evolution.

To investigate the origin of these correlation effects between H and He in $\alpha$-Fe is the purpose 
of this paper. We provide an atomistic description of an extensive set of small clusters of He, H and 
vacancies (V$_n$He$_m$H$_p$, with n,m,p up to 6), presenting first principles calculations to understand 
the interaction between He and H in iron. By comparing the energetics of these clusters we argue that 
the direct interaction between He and H is marginal (mostly repulsive) and that the ostensible strong 
attraction reported by others~\cite{Tian2010} is due to the preferential clustering of helium or hydrogen 
to iron vacancies and not to the direct interaction between helium and hydrogen. 
Additional kinetic simulations based on a Rate Theory approach are presented to illustrate the expected 
effects due to the combined presence of H and He on their evolution under typical irradiation conditions 
in a fusion environment.

Calculations are performed using the spin-polarized generalized gradient approximation 
(GGA)~\cite{PBE} as implemented in the SIESTA code.~\cite{siesta} 
Norm-conserving pseudopotentials~\cite{pseudos} with cutoff radii of 1.11, 1.21, 0.71, and 
1.06 \AA{} are used for Fe $4s$, $4p$, $3d$ and $4f$, and a cutoff radius of 0.45 \AA{} for 
the partial-core correction.  H and He pseudopotentials are built with cutoff radii of 0.66 
and 0.69 \AA{}, respectively. Valence electrons are described with a linear combination of 
strictly localized numerical atomic orbitals that vanish outside certain cutoff radii 
($r_c$, in \AA{}).  Triple-$\zeta$ basis was used for Fe $4s$ ($r_c=4.84$) and $3d$ ($r_c=3.06$) 
orbitals and double-$\zeta$ for Fe $4p$ ($r_c=3.53$), H and He $1s$ ($r_c=2.63$).~\cite{GarciaSuarez2004}
Additional (single-$\zeta$) polarization orbitals were considered for the H and He electrons.
Supercell calculations were first performed on ferromagnetic Fe$_{54}$ boxes to explore the energetics of the 
different defect configurations, and then repeated for larger Fe$_{128}$ cells.
A real space mesh of 0.079 \AA{} was used, and at least 4$\times$4$\times$4 k-point grid samplings 
for full relaxation of the atomic positions (forces smaller than 0.04 eV/\AA{}), keeping the 
volume of the cell fixed.

The zero-point energy (ZPE) is known to be important for H defects,~\cite{Tateyama2003} and
can be computed from the vibrational frequencies of the normal modes using 
ZPE=$\frac{1}{2}\sum_{\nu}\hbar\nu$, where the frequencies $\nu$ are obtained assuming 
the Fe atoms are fixed at their equilibrium positions (Fe mass is much larger than H or He 
masses), and only the light atoms vibrate (Einstein approximation). The ZPE of $\frac{1}{2}$H$_2$
is 0.13 eV, in good agreement with previous results,~\cite{Jiang2004,Counts2010} and comparable to
the average ZPE for VH$_n$ complexes (0.14 eV/H) reported by Tateyama and Ohno.~\cite{Tateyama2003}
Similar values are obtained here for interstitial H in bulk Fe (0.15 eV and 0.19 eV for octahedral 
and tetrahedral sites), and close to a substitutional He (0.15 eV). This homogeneity of the ZPE 
for H in different configurations allows to take an average contribution for ZPE of $\sim0.15$eV per 
H atom for each complex cluster.~\cite{Tateyama2003}

\begin{figure}
\includegraphics[width=0.45\textwidth]{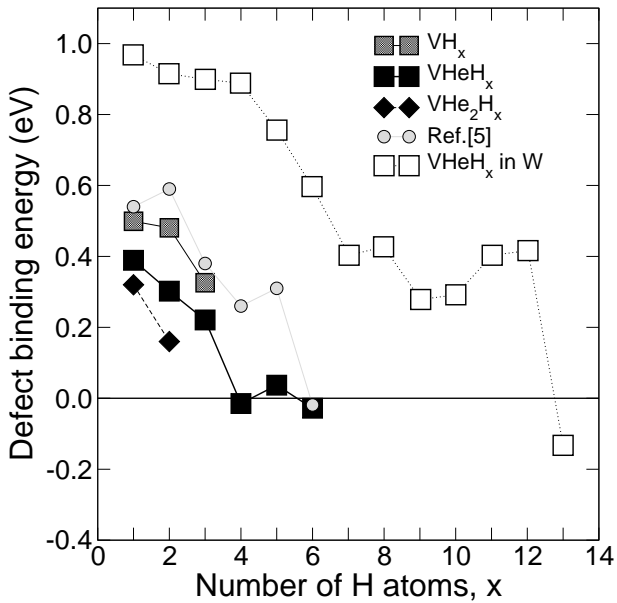}
\caption{\label{binding-H} Binding energies of interstitial H to a VH$_x$ (gray squares), 
VHeH$_x$ (black squares), and VHe$_2$H$_x$ (black diamonds) defect clusters. Also shown the
binding energies for VH$_x$ reported by Tateyama {\it et. al.} ref.[\onlinecite{Tateyama2003}]
(light gray circles) and the binding energies to a single substitutional He in tungsten 
(empty squares)\cite{Jiang2010}. The horizontal axis refers to the number of H atoms in 
the initial cluster, $x$.}
\end{figure}
One approach to quantify the interaction between H and He is through their {\it binding energies}.
The binding energy of hydrogen to a defect cluster, V$_n$He$_m$H$_{p}$, is defined as 
the energy gained by trapping an additional H into that cluster to form V$_n$He$_m$H$_{p+1}$.  
A positive energy indicates an attraction of H to the defect cluster. 
Based on this representation Jiang and collaborators\cite{Jiang2010} suggest a strong hydrogen 
trapping at helium in W (shown in figure \ref{binding-H} with empty squares), with up to twelve H 
atoms clustering around one single substitutional He. 
Trapping at V$_1$He$_1$ in iron is notably weaker, with a binding energy of 0.2-0.4 eV with a maximum 
of five hydrogens around a helium atom. 
We found that the binding of H to a single vacancy site (no He) is also positive, in good agreement 
with the results reported by Tateyama and coworkers~\cite{Tateyama2003} (light gray circles) that 
showed that a vacancy can accommodate no more than six hydrogen atoms.
Notice however the decrease of $\sim$0.1 eV with respect to the H binding energy to a single vacancy 
(gray squares) observed when a helium atom is placed at the vacancy (black squares).

\begin{figure}
\includegraphics[width=0.40\textwidth]{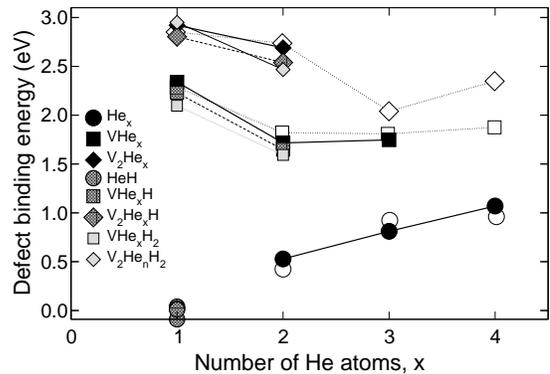}
\caption{\label{binding-He} Binding energies of interstitial He to V$_n$He$_m$H$_p$ complexes. 
Empty symbols show the binding to interstitials He (circles), VHe$_x$ (squares) and V$_2$He$_x$ 
(diamonds) reported in ref.[\onlinecite{Fu2005_2}]. Our results are shown as solid black symbols.
Dark (light) gray symbols correspond to the binding energies when one (two) H atoms sit close to the 
vacancy position.  The horizontal axis refers to the number of He atoms in the initial cluster, $x$.}
\end{figure}

We can similarly plot the binding energy of an interstitial He to form a V$_n$He$_{m+1}$H$_p$ cluster 
(figure \ref{binding-He}). In agreement with previous reports,~\cite{Fu2005_2} there is a strong
self-trapping of He atoms in bulk He (circles in the figure) and an even larger trapping at 
vacancy complexes. However, the attractive interaction is again decreased (by tenths of eV) 
when hydrogen atoms are present. In other words, although interstitial He lowers its energy 
by approaching a vacancy site, the presence of H at the vacancy inhibits He trapping.
Further evidence that He and H do not interact strongly is given by the small (even negative)
binding energies of He to interstitial H (gray circles).
In what follows, we will argue that the positive binding energies for H at V$_n$He$_m$ clusters 
(Fig.~\ref{binding-H}) or He at V$_n$H$_p$ clusters (Fig.~\ref{binding-He}) are not due to the 
presence of He or H atoms, but to the vacancy itself.

\begin{figure*}
\includegraphics[width=0.75\textwidth]{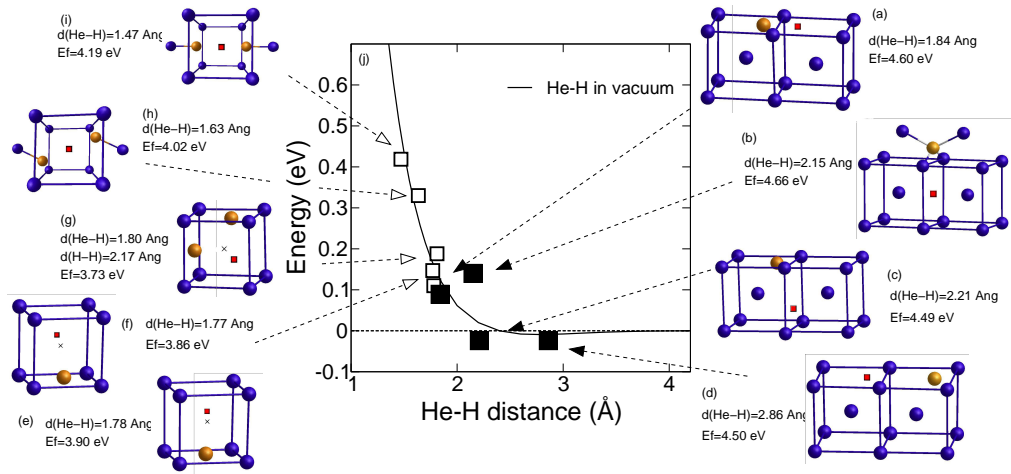}
\caption{\label{defects} (color-online) Atomic configurations for He and H interstitials (a-d)
V$_1$He$_1$H$_1$ (e-f) and V$_1$He$_1$H$_2$ defects (g-i) in bulk Fe.  
Blue (dark gray) spheres, orange (light gray) and red square correspond to Fe, H, and He atoms. 
Vacancy sites are marked with a small cross. The interatomic distance and defect formation energies 
are also shown.
Central panel (j) shows the interaction energy for He-H pairs in vacuum (solid line) as a function 
of the interatomic distance, compared to the energies for defects in bulk iron. 
Solid squares are obtained from interstitial pairs and empty symbols are obtained from V$_1$He$_1$H$_1$ 
and V$_1$He$_1$H$_2$ clusters, as highlighted by dashed arrows.
}
\end{figure*}

The interatomic potential for He-H in iron can be determined from the dependence of their 
binding energy on the He-H distance.
We can do this for a relevant subset of the defect configurations studied in this work, which is shown in 
Fig.~\ref{defects} together with their corresponding He-H interatomic distances and {\it formation energies}.
The defect formation energies are defined relative to the chemical potentials taken from bulk 
$\alpha$-Fe, molecular hydrogen, and the isolated He atom. 
At the right panels, (a) to (d), different configurations for interstitials of He and H are 
presented (V$_0$He$_1$H$_1$ clusters). 
Left panels, (e) to (i), show several V$_1$He$_1$H$_1$ and V$_1$He$_1$H$_2$ clusters, and
illustrate how hydrogen affects the stability of VHe, displacing He from the substitutional site,
and moving itself from the preferential tetrahedral occupation of interstitial H, towards the octahedral site.

Central panel in Fig.~\ref{defects}(j) shows the interatomic potential (binding energies) 
for He-H pairs computed in bulk $\alpha$-Fe. 
Filled squares are obtained from the difference between the energy of a HeH complex, and 
the sum of the energy of interstitial hellium and interstitial hydrogen.
%
On the other hand, empty squares are defined from the decrease (per H atom) on the binding energy
of one He atom to a vacancy due to the presence of hydrogen atoms at the vacancy (difference between
black, dark gray and light gray squares at x=1 in Fig.~\ref{binding-He}, divided by the number 
of H atoms). 
This rough approximation follows with reasonable accuracy the interatomic potential for He and H 
computed in vacuum (solid line). It shows that the binding of He and H in bulk iron is only 
marginal, and could be modeled with the simple two-body potential of isolated He and H atoms.
Although there is a small attractive interaction for interstitials in configurations (c) and (d), 
for which the respective He-H distances are 2.21\AA\ and 2.86\AA\, the interaction is clearly 
repulsive for interatomic distances smaller than $\sim$2\AA, which is the case for all the 
studied clusters where He and H are trapped in a vacancy.

These \textit{ab initio} calculations show that, although the direct interaction between H and He is weak, 
they both bind strongly to vacancies, which could be at the origin of the experimentally observed synergic 
effects.~\cite{Tanaka2004_1,Fukushima85,Schroeder91} To investigate this point, we simulated the 
evolution of H and He in $\alpha$-Fe under irradiation, in conditions similar to those expected in fusion 
environment.  Nuclear transmutation continously generates H and He that evolve
in the presence of a large amount of defects created by atomic displacements due to collisions between 
energetic neutrons and lattice atoms.  
Fusion conditions expected in a DEMO-HCLL-4000MW reactor configuration were considered.
A neutron flux of 1.52$\times 10^{15}$ /cm$^{2}$/s was calculated\cite{Jordanova_1} and 
transmutation rates of 1.75$\times 10^{12}$ and 1.1$\times 10^{13}$ /cm$^3$/s were obtained 
for He and H in Fe using the nuclear inventory code ACAB\cite{ACAB} and the EAF 2007 library\cite{EAF}.
%
%
The generation rate of atomic displacements (Frenkel pairs) corresponding to the neutron spectrum was 
calculated using the methodology presented in Ref. \onlinecite{Mota_1} and was found to be about 
1.82$\times 10^{16}$ /cm$^3$/s. 
To simulate the evolution of H, He and defects we expanded the Rate Theory model developed by Ortiz 
\textit{et al}\cite{Ortiz2009_1} to predict the diffusion and clustering of He in $\alpha$-Fe in 
the presence of impurities. The various V$_n$He$_m$H$_p$ clusters studied from \textit{ab initio} 
in the present work were added to the model, with their corresponding binding energies and reaction 
rates, calculated similarly to what is done in Ref. \onlinecite{Ortiz2009_1}. 
It is important to note that the growth of large bubbles containing both H and He is not yet 
implemented in our model, since their energetics is not known, and only the limited number of 
small V$_n$He$_m$H$_p$ clusters is included.

\begin{figure}
\includegraphics[width=0.40\textwidth]{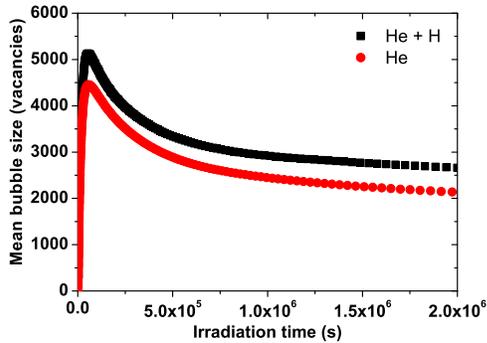}
\caption{\label{kinetic} Evolution of the mean bubble size in irradiated Fe for a system 
where only He is present (red dots) and for a system where H and He evolve simultaneously (black squares).}
\end{figure}

For our purpose, we considered a temperature of 450$^\circ$C, which falls in the range of operating 
temperature (350-500$^\circ$C) expected for DEMO reactor\cite{Zinkle2011_1}, depending on the cooling options. 
Fig. \ref{kinetic} reports the evolution of the mean bubble size as a function of irradiation time for the 
H-He system described above. For comparison, the evolution of the mean bubble size calculated for a system 
where only He is generated (no H) is also shown. Our simulations clearly predict that a larger mean bubble 
size is expected when H and He evolve simultaneously. 
This result follows the trend of the experimental measurements obtained in FeCr alloys by Tanaka \textit{et. al.},
~\cite{Tanaka2004_1} who observed larger bubbles when H is present. 
It is important to note here that only the mean size of He bubbles is shown in Fig.~\ref{kinetic}.
The contribution of the small V$_n$He$_m$H$_p$ clusters to the mean bubble size is minor. 
Hence, the larger mean size of bubbles that is predicted by our simulations when H is present 
is not due to large bubbles containing H but to a synergic effect between H and He. 
Remarkably, a significant effect is predicted though only few V$_n$He$_m$H$_p$ 
configurations are included in the kinetic model.


Our simulations confirm that significant synergic effects between H and He could be expected in 
some cases, though their direct interaction might be weak. the apparent strong positive interaction 
reported in the literature is due to the much larger V-H and V-He attraction that conceals the 
repulsive interaction between He and H.  They both tend to be trapped by voids, and this, 
rather than the He-H interaction, could explain the reduction of hydrogen permeation, and the larger 
cavity sizes in ferritic alloys simultaneously irradiated with H and He.~\cite{Schroeder91,Tanaka2004_1,Fukushima85}
This has important implications for modeling damage evolution in materials for fusion, that until
now have mainly considered the role of isolated He or H aggregates, and not their coexistence. 
To understand in which conditions synergetic effects could be expected, further work will be needed
on the energetics of larger clusters or bubbles containing both He and H in steels, and
on the effect of solutes and grain boundaries where bubbles tend to nucleate.

The authors acknowledge financial support by the European fusion materials modeling programme (EFDA) 
and the Spanish MCINN (FIS2009-12721-C04-01 and CSD2007-00041). CJO also acknowledges F. Mota for 
neutron transport calculations.

\bibliography{biblio}

\end{document}